\documentclass[preprintnumbers,amsmath,amssymb,showpacs]{revtex4}
\usepackage{graphicx}% Include figure files
\usepackage{dcolumn}% Align table columns on decimal point
\usepackage{bm}% bold math
\usepackage{booktabs}
\usepackage{CJK}
\usepackage{subfigure}
\usepackage{qcircuit}
\usepackage{color}
\usepackage{braket}
\usepackage{listings}
\usepackage{soul}
\usepackage{xcolor}
\soulregister\cite7 % Õë¶Ô\cite ÃüÁî
\soulregister\ref7 % Õë¶Ô\ref ÃüÁî
\usepackage{color, soul} %ÓÃcolor, ºÍ soul °ü
\setstcolor{yellow}

\begin{document}

\title{Quantum teleportation based on the elegant joint measurement}

\author{Dong Ding$^{1}$}
\author{Ming-Xing Yu$^{1}$}
\author{Ying-Qiu He$^{1}$}
\email{heyq@ncist.edu.cn}
\author{Hao-Sen Ji$^{1}$}
\author{Ting Gao$^{2}$}
\email{gaoting@hebtu.edu.cn}
\author{Feng-Li Yan$^{3}$}
\email{flyan@hebtu.edu.cn}

\affiliation {
$^1$ College of Science, North China Institute of Science and Technology, Beijing 101601, China\\
$^2$ School of Mathematical Sciences, Hebei Normal University, Shijiazhuang 050024, China\\
$^3$ College of Physics, Hebei Normal University, Shijiazhuang 050024, China}
\date{\today}

\begin{abstract}

As a generalization of the well-known Bell state measurement (BSM), the elegant joint measurement (EJM) is a kind of novel two-qubit joint measurement, parameterized by a subtle phase factor $\theta \in [0,\pi/2]$.
We explore quantum teleportation based on the EJM, inspired by Gisin's idea that quantum entanglement not only provides quantum channel and also quantum joint measurement for quantum teleportation.
It is a probabilistic teleportation caused by undesired nonunitary quantum evolution.
There are two interesting features in the present scenario. First, it goes beyond the conventional teleportation scenario, which can be included in the present scenario.
Second, different from the BSM being single input and four outcomes, it can provide an adjustable input setting or even multiple measurement settings for the sender (or the controller).
Moreover, we show in detail the feasible quantum circuits to realize the present scenario, where a few unitary operations and a nonunitary quantum gate are being utilized.

\end{abstract}


%Keywords: quantum teleportation; quantum join measurement; nonunitary quantum gate

%03.65.Ud:Entanglement and quantum nonlocality; 03.67.-a:Quantum information;
%42.50.-p:Quantum optics; 03.67.Lx Quantum computation architectures and implementations
%03.67.Hk Quantum communication; 03.67.Mn Entanglement measures, witnesses, and other characterizations
%
\maketitle

\section{Introduction}

In 1993, Bennett \emph{et al.} \cite{Bennett1993} proposed a method to transfer an unknown quantum state to another distant system, i.e. quantum teleportation. It has attracted much attention over the last 30 years, mainly because of its potential for secure communication \cite{NC2000,Li2023}. These include open-destination teleportation \cite{KB1998}, probabilistic teleportation \cite{PQT-GUO2000}, port-based teleportation \cite{port-based-QT2008,port-based-QT2023}, high-dimensional teleportation \cite{Luo2019}, two-copy quantum teleportation \cite{Two-copy-QT2020}, and experimental realization in various physical systems \cite{E-QT-BPM1997,E-QT-NKL1998,E-QT-atomic2004,E-QT-solid-state2013,E-QT-Optomechanical2021}.

The basic procedure for teleporting a quantum state is three-fold: (i) specify an entangled resource shared between two distant communication parties, as a quantum teleportation channel; (ii) the sender performs a local joint measurement on his (or her) subsystem and then tells the result to the receiver through a classical channel; (iii) the receiver performs a specific unitary operation based on the measurement result to recover the teleported quantum state.
In general, quantum entanglement \cite{Quantum-entanglement,Entanglement-detection2009} is exploited twice in teleportation, providing quantum teleportation channel the first time, and the eigenvectors of joint measurement the second time \cite{Gisin-EJM2019}.
On the other hand, in his original paper, Gisin \cite{Gisin-EJM2019} suggested that quantum entanglement enables entirely new kinds of joint measurements, and then proposed a two-qubit joint measurement, named elegant joint measurement (EJM). The reduced states (described by tracing out either of the parties) of Bell state measurement (BSM) bases are the completely mixed state $\rho=I/2$, corresponding to zero Bloch vector; on the contrary, the reduced states of EJM bases, are non-zero Bloch vectors, which have elegant symmetry property, and thus named the elegant joint measurement.
More recently, Tavakoli \emph{et al.} \cite{TGB-EJM2021} extended Gisin's original EJM to a one-parameter family, and then investigated quantum violations of bilocality inequalities in quantum network featuring independent sources \cite{Tavakoli-network2022,BGT2021IBM}.
By now, another interesting application of EJM is to experimentally realize entanglement swapping with hyperentanglement photons \cite{PanPRL-EJM2022}.

Inspired by Gisin's work \cite{Gisin-EJM2019}, in this paper, we would like to develop quantum teleportation based on the EJM.
In the process of performing local joint measurement, we shall carry out the EJM instead of the conventional BSM.
Each unitary operation on the receiver's system used to recover the teleported state is, accordingly, replaced by a nonunitary one.
We instead adopt the probabilistic nonunitary gate \cite{TU-Nonunitary2005} consisting of two single qubit unitary gates and a controlled unitary gate with the aid of an ancilla qubit followed by projective measurement.
It is a probabilistic teleportation scenario and thus we calculate the probability of success for teleporting an arbitrary single qubit state.
Furthermore, we provide quantum circuits for implementing the present scenario, involving the initial state preparation, the EJM on the sender's system and the corresponding nonunitary quantum evolution by the receiver.

\section{The elegant joint measurement}

In this section, we review the results on the EJM \cite{Gisin-EJM2019,TGB-EJM2021,Tavakoli-network2022}.
An EJM, parameterized by $\theta \in [0,\pi/2]$, is a projection onto the four basis states in two-qubit Hilbert space, i.e.
\begin{eqnarray}
\left |e_{00}  \right \rangle =
\frac{1}{2} ( \text{e}^{-\frac{\text{i}\pi }{4}}, r_{-}^{\theta }, r_{+}^{\theta }, \text{e}^{-\frac{3\text{i}\pi }{4}} )^{\dag},
\end{eqnarray}
\begin{eqnarray}
\left |e_{01}  \right \rangle =
\frac{1}{2} ( \text{e}^{\frac{3\text{i}\pi }{4}}, r_{-}^{\theta }, r_{+}^{\theta }, \text{e}^{\frac{\text{i}\pi }{4}}  )^{\dag},
\end{eqnarray}
\begin{eqnarray}
\left |e_{10}  \right \rangle =
\frac{1}{2} ( \text{e}^{\frac{\text{i}\pi }{4}}, -r_{+}^{\theta }, -r_{-}^{\theta },
\text{e}^{\frac{3 \text{i}\pi }{4}} )^{\dag},
\end{eqnarray}
\begin{eqnarray}
\left |e_{11}  \right \rangle =
\frac{1}{2} ( \text{e}^{-\frac{3\text{i}\pi }{4}}, -r_{+}^{\theta }, -r_{-}^{\theta}, \text{e}^{-\frac{\text{i}\pi }{4}} )^{\dag},
\end{eqnarray}
where
$r_{\pm }^{\theta } =(1\pm \text{e}^{-\text{i}\theta}) /\sqrt{2}$.
There are two elegant properties of these states: (i) all elements are equally entangled, and (ii) the two sets of four reduced states, corresponding to either qubit being traced out, form two regular tetrahedrons of radius $\sqrt{3}/2 \cos\theta$ inside the Bloch sphere, respectively.
Notice that for $\theta=0$ it matches the largest local tetrahedron of radius $\sqrt{3}/2$, while for $\theta=\pi/2$ it has the smallest local tetrahedron of radius zero and the EJM is equivalent to the BSM up to local unitary transformations.

\section{Quantum teleportation based on the EJM}

\subsection{Scenario of teleportation}

We will simplify the discussion by only considering teleporting a single qubit state using the singlet state as quantum channel in the text.
Assume that the sender Alice wishes to teleport an unknown single qubit state
\begin{eqnarray} \label{unknown-state}
\left | \psi_{0}  \right \rangle_{1} =\alpha \left | 0  \right \rangle_{1} + \beta \left | 1  \right \rangle_{1},
\left | \alpha  \right |^{2} + \left | \beta  \right |^{2}=1,
\end{eqnarray}
to the receiver Bob, where Alice and Bob share an entangled pair of qubits in the singlet state
\begin{eqnarray}
\left | \Psi^{-}  \right \rangle_{23} =\frac{1}{\sqrt{2} } \left ( \left |01\right \rangle_{23}-\left |10\right \rangle_{23} \right ),
\end{eqnarray}
as the quantum channel. The first two qubits (1 and 2) belong to Alice, and the third one to Bob.
So the combined system including three qubits is initially described by the state
\begin{eqnarray}
\left | \Phi   \right \rangle _{123}=\left | \psi_{0}  \right \rangle_{1}\otimes \left | \Psi^{-}  \right \rangle_{23}=\frac{1}{\sqrt{2} }\left ( \alpha \left | 001  \right \rangle_{123}-\alpha \left | 010  \right \rangle_{123}+\beta \left | 101  \right \rangle_{123}-\beta \left | 110  \right \rangle_{123}  \right ).
\end{eqnarray}

Now, Alice performs an EJM on qubits 1 and 2 described by $\{|e_{i}\rangle\langle e_{i}|\}, i=00,01,10,11$, where the parameter $\theta$ is completely specified before the scenario begins. It can be a single measurement setting or be multiple measurement settings \cite{Bell-nonlocality2014,GYE-PRL2014,HDYG2015EPL,DHYG-JPA2020} (e.g. binary $\theta_{0}=0$ and $\theta_{1}=\pi/2$, for simplicity), eventually decided by the specific quantum teleportion tasks.

As a result, if Alice obtains the result $\left |e_{00}  \right \rangle$ then Bob's system will be in the (un-normalized) state
\begin{eqnarray}\label{psi00}
|\psi_{00}\rangle = -r_{-}^{\theta} \alpha \left | 0  \right \rangle +\text{e}^{-\frac{\text{i}\pi}{4}}\alpha \left | 1  \right \rangle -\text{e}^{-\frac{3 \text{i}\pi}{4}}\beta\left | 0  \right \rangle + r_{+}^{\theta} \beta \left | 1  \right \rangle;
\end{eqnarray}
similarly, for Alice's results $\left |e_{01}  \right \rangle$, $\left |e_{10}  \right \rangle$ and $\left |e_{11}  \right \rangle$, the resulting states of Bob's qubit 3 will be respectively
\begin{eqnarray}
|\psi_{01}\rangle = -r_{-}^{\theta} \alpha \left | 0  \right \rangle +\text{e}^{\frac{3 \text{i}\pi}{4}}\alpha \left | 1  \right \rangle -\text{e}^{\frac{\text{i}\pi}{4} }\beta\left | 0  \right \rangle + r_{+}^{\theta} \beta \left | 1  \right \rangle,
\end{eqnarray}
\begin{eqnarray}\label{psi10}
|\psi_{10}\rangle = r_{+}^{\theta } \alpha \left | 0  \right \rangle +\text{e}^{\frac{\text{i}\pi}{4}}\alpha \left | 1  \right \rangle -\text{e}^{\frac{3 \text{i}\pi }{4} }\beta\left | 0  \right \rangle - r_{-}^{\theta} \beta \left | 1  \right \rangle,
\end{eqnarray}
and
\begin{eqnarray}
|\psi_{11}\rangle = r_{+}^{\theta } \alpha \left | 0  \right \rangle +\text{e}^{-\frac{3 \text{i}\pi}{4}}\alpha \left | 1  \right \rangle -\text{e}^{-\frac{\text{i}\pi}{4}}\beta\left | 0  \right \rangle - r_{-}^{\theta} \beta \left | 1  \right \rangle.
\end{eqnarray}
Let $N_{i}=1/\sqrt{\langle\psi_{i}|\psi_{i}\rangle}$ ($i=00,01,10,11$) be the normalization factors required to maintain $\text{tr}(N_{i}^{2}|\psi_{i}\rangle\langle\psi_{i}|)=1$.
The probability for obtaining $|e_{i} \rangle$ is $p_{i}=\langle\psi_{i}|\psi_{i}\rangle/8=1/(8N_{i}^{2})$.

Then, Alice tells Bob the result of EJM over a classical communication channel.
Depending on Alice's information, the next work of Bob is to accurately recover the original quantum state, i.e. finding $A_{i}|\psi_{i}\rangle = |\psi_{0}\rangle, i=00,01,10,11$.

A straightforward calculation shows that
\begin{eqnarray}
A_{00}=\frac{\sqrt{2}}{3-\text{e}^{-2\text{i}\theta}}
\left(\begin{array}{cc}
 -1-\text{e}^{-\text{i}\theta }  & 1+\text{i}
\\1-\text{i} & 1-\text{e}^{-\text{i}\theta }
\end{array}\right),
\end{eqnarray}
\begin{eqnarray}
A_{01}=\frac{\sqrt{2}}{3-\text{e}^{-2\text{i}\theta}}
\left(\begin{array}{cc}
 -1-\text{e}^{-\text{i}\theta }  & -1-\text{i}
\\-1+\text{i} & 1-\text{e}^{-\text{i}\theta }
\end{array}\right),
\end{eqnarray}
\begin{eqnarray}
A_{10}=\frac{\sqrt{2}}{3-\text{e}^{-2\text{i}\theta}}
\left(\begin{array}{cc}
 1-\text{e}^{-\text{i}\theta }  & 1-\text{i}
\\1+\text{i} & -1-\text{e}^{-\text{i}\theta }
\end{array}\right)
\end{eqnarray}
and
\begin{eqnarray}
A_{11}=\frac{\sqrt{2}}{3-\text{e}^{-2\text{i}\theta}}
\left(\begin{array}{cc}
 1-\text{e}^{-\text{i}\theta }  & -1+\text{i}
\\-1-\text{i} & -1-\text{e}^{-\text{i}\theta }
\end{array}\right),
\end{eqnarray}
corresponding to the resulting states $|\psi_{00}\rangle, |\psi_{01}\rangle, |\psi_{10}\rangle$ and $|\psi_{11}\rangle$ to $|\psi_{0}\rangle$, respectively.

By this, as an extension of the conventional quantum teleportation by means of the BSM, the present scenario based on the EJM enables us to teleport an unknown quantum state from one location to another.
On the other hand, it is clear, however, that these evolution matrices are nonunitary, i.e. $A_{i}^{\dag}A_{i}=A_{i}A_{i}^{\dag} \neq 1, i=00,01,10,11$, except for $\theta=\pi/2$, which is equivalent to the BSM up to local unitary transformations.

Quantum information processing is usually described by unitary evolution, since the evolution of a closed quantum system obeys the unitarity constraint. So, how to implement these nonunitary evolutions on Bob's qubit is the key ingredient for teleportation in the present scenario.

\subsection{Characterization of the nonunitary quantum evolutions}

Quantum mechanically, it is impossible to perform deterministic nonunitary quantum gate but is possible to realize a probabilistic one by considering ancilla qubit followed by projective measurement \cite{TU-Nonunitary2005,SAMT-Nonunitary2016}.

In general, a quantum measurement is described by a set of measurement operators $\{M_{m}\}$,  satisfying the completeness equation $\sum_{m}M_{m}^{\dag}M_{m}=I$.
If one performs a measurement on a quantum system $|\psi\rangle$, then it would be possible to
obtain the result $m$ with probability $p(m)=\langle\psi|M_{m}^{\dag}M_{m}|\psi\rangle$, leaving the postmeasurement state $M_{m}|\psi\rangle/\sqrt{p(m)}$.
By this representation, we next deal with these nonunitary evolution matrices $A_{i}$s, $i=00,01,10,11$, as follows.

Let
\begin{eqnarray}
M_{0,i}=c_{i}A_{i}, ~~ M_{1,i}=\sqrt{I-(M_{0,i}^{\dag}M_{0,i})},
\end{eqnarray}
where each $c_{i}$ is a complex coefficient.
If we perform the measurement $\{M_{0,i}, M_{1,i}\}$ on Bob's system (qubit 3), then each nonunitary matrix $A_{i}$ can be described by the following quantum operation
\begin{eqnarray}
\mathcal{N}(|\psi_{i}\rangle\langle\psi_{i}|) = M_{0,i}|\psi_{i}\rangle\langle\psi_{i}|M_{0,i}^{\dag}
\end{eqnarray}
with success probability
\begin{eqnarray} \label{probability-c}
p(|\psi_{i}\rangle;c_{i})=\langle\psi_{i}|M_{0,i}^{\dag}M_{0,i}|\psi_{i}\rangle
=|c_{i}|^{2}\langle\psi_{i}|A_{i}^{\dag}A_{i}|\psi_{i}\rangle.
\end{eqnarray}
The postmeasurement state is accordingly written as
\begin{eqnarray}
\frac{M_{0,i}|\psi_{i}\rangle}{\sqrt{p(|\psi_{i}\rangle;c_{i})}}
=\frac{c_{i}}{|c_{i}|}\frac{A_{i}|\psi_{i}\rangle}{\sqrt{\langle\psi_{i}|A_{i}^{\dag}A_{i}|\psi_{i}\rangle}}.
\end{eqnarray}
It is exactly the state $|\psi_{0}\rangle$ being teleported up to an unobservable global phase factor $c_{i}/|c_{i}|$, where the coefficient $c_{i}$ does not affect the postmeasurement state while it will affect the probability of success.

We now proceed to describe this procedure in detail.
To do this, we first make the singular value decomposition \cite{NC2000} of $A_{i}$, i.e.
\begin{eqnarray}
A_{i}=U_{i}D_{i}V_{i}^{\dag},
\end{eqnarray}
where $U_{i}$ and $V_{i}$ are unitary matrices (operators) and $D_{i}$ is a diagonal matrix.
A straightforward but lengthy calculation shows that
\begin{eqnarray}
A_{00}=
\left(\begin{array}{cc}
\frac{-(1+\text{e}^{-\text{i}\theta}) (\sqrt{3}+1) - 2}{a_{+}d_{+}^{\theta}}  &
\frac{(1+\text{e}^{-\text{i}\theta}) (\sqrt{3}-1) - 2}{a_{-}d_{-}^{\theta}}\\
\frac{\sqrt{2}\text{e}^{-\frac{\text{i}\pi}{4}}(\sqrt{3}+\text{e}^{-\text{i}\theta})}{a_{+}d_{+}^{\theta}} &
\frac{-\sqrt{2}\text{e}^{-\frac{\text{i}\pi}{4}}(\sqrt{3}-\text{e}^{-\text{i}\theta})}{a_{-}d_{-}^{\theta}}
\end{array}\right)
\left(\begin{array}{cc}
 d_{+}^{\theta}  & 0\\
 0  & ~d_{-}^{\theta}
\end{array}\right)
\left(\begin{array}{cc}
 \frac{\sqrt{3}+1}{a_{+}}   & \frac{\sqrt{2}\text{e}^{-\frac{3 \text{i}\pi}{4}}}{a_{+}}\\
-\frac{\sqrt{3}-1}{a_{-}}   & \frac{\sqrt{2}\text{e}^{-\frac{3 \text{i}\pi}{4}}}{a_{-}}
\end{array}\right),
\end{eqnarray}

\begin{eqnarray}
A_{01}=
\left(\begin{array}{cc}
\frac{-(1+\text{e}^{-\text{i}\theta}) (\sqrt{3}+1) - 2}{a_{+}d_{+}^{\theta}}  &
\frac{-(1+\text{e}^{-\text{i}\theta}) (\sqrt{3}-1) + 2}{a_{-}d_{-}^{\theta}}\\
\frac{\sqrt{2}\text{e}^{\frac{3\text{i}\pi}{4}}(\sqrt{3}+\text{e}^{-\text{i}\theta})}{a_{+}d_{+}^{\theta}} &
\frac{\sqrt{2}\text{e}^{\frac{3\text{i}\pi}{4}}(\sqrt{3}-\text{e}^{-\text{i}\theta})}{a_{-}d_{-}^{\theta}}
\end{array}\right)
\left(\begin{array}{cc}
 d_{+}^{\theta}  & 0\\
 0  & ~d_{-}^{\theta}
\end{array}\right)
\left(\begin{array}{cc}
\frac{\sqrt{3}+1}{a_{+}}   &  \frac{\sqrt{2}\text{e}^{\frac{\text{i}\pi}{4}}}{a_{+}}\\
\frac{\sqrt{3}-1}{a_{-}}   & -\frac{\sqrt{2}\text{e}^{\frac{\text{i}\pi}{4}}}{a_{-}}
\end{array}\right),
\end{eqnarray}

\begin{eqnarray}
A_{10}=
\left(\begin{array}{cc}
\frac{-(1-\text{e}^{-\text{i}\theta}) (\sqrt{3}-1) +2}{a_{-}d_{+}^{\theta}}  &
\frac{(1-\text{e}^{-\text{i}\theta}) (\sqrt{3}+1) +2}{a_{+}d_{-}^{\theta}}\\
\frac{-\sqrt{2}\text{e}^{\frac{\text{i}\pi}{4}}(\sqrt{3}+\text{e}^{-\text{i}\theta})}{a_{-}d_{+}^{\theta}} &
\frac{\sqrt{2}\text{e}^{\frac{\text{i}\pi}{4}}(\sqrt{3}-\text{e}^{-\text{i}\theta})}{a_{+}d_{-}^{\theta}}
\end{array}\right)
\left(\begin{array}{cc}
 d_{+}^{\theta}  & 0\\
 0  & ~d_{-}^{\theta}
\end{array}\right)
\left(\begin{array}{cc}
 -\frac{\sqrt{3}-1}{a_{-}}  & -\frac{\sqrt{2}\text{e}^{\frac{3 \text{i}\pi}{4}}}{a_{-}}\\
 \frac{\sqrt{3}+1}{a_{+}}   & -\frac{\sqrt{2}\text{e}^{\frac{3 \text{i}\pi}{4}}}{a_{+}}
\end{array}\right)
\end{eqnarray}
and
\begin{eqnarray}
A_{11}=
\left(\begin{array}{cc}
\frac{(1-\text{e}^{-\text{i}\theta}) (\sqrt{3}-1) -2}{a_{-}d_{+}^{\theta}}  &
\frac{(1-\text{e}^{-\text{i}\theta}) (\sqrt{3}+1) +2}{a_{+}d_{-}^{\theta}}\\
\frac{\sqrt{2}\text{e}^{-\frac{3\text{i}\pi}{4}}(\sqrt{3}+\text{e}^{-\text{i}\theta})}{a_{-}d_{+}^{\theta}} &
\frac{\sqrt{2}\text{e}^{-\frac{3\text{i}\pi}{4}}(\sqrt{3}-\text{e}^{-\text{i}\theta})}{a_{+}d_{-}^{\theta}}
\end{array}\right)
\left(\begin{array}{cc}
 d_{+}^{\theta}  & 0\\
 0  & ~d_{-}^{\theta}
\end{array}\right)
\left(\begin{array}{cc}
\frac{\sqrt{3}-1}{a_{-}}  &  \frac{\sqrt{2}\text{e}^{-\frac{\text{i}\pi}{4}}}{a_{-}}\\
\frac{\sqrt{3}+1}{a_{+}}  & -\frac{\sqrt{2}\text{e}^{-\frac{\text{i}\pi}{4}}}{a_{+}}
\end{array}\right),
\end{eqnarray}
up to the complex coefficient $\sqrt{2}/(3-\text{e}^{-2\text{i}\theta})$, where $a_{\pm}=\sqrt{6\pm2\sqrt{3}}$ and $d_{\pm}^{\theta}=\sqrt{4 \pm 2\sqrt{3}\cos \theta}$.

By the singular value decomposition, it is not difficult to see that all of the matrices $A_{i}$s have the same diagonal matrix $D_{i} \equiv D=\text{diag}(d_{+}^{\theta}, d_{-}^{\theta})$, being nonunitary matrix except for $\theta=\pi/2$.
Since an arbitrary single qubit unitary operator $U$ can always be written as several rotations and a global phase, these unitary matrices can be further decomposed as $U=\text{e}^{\text{i}\alpha}R_{z}( \beta ) R_{y}( \gamma)R_{z}( \delta)$, for example. Thus, the main workhorse here is how to realize this nonunitary diagonal matrix $D$.

Consider a single qubit gate  \cite{TU-Nonunitary2005}
\begin{eqnarray}
N(d)=
\left(\begin{array}{cc}
1  &   0\\
0  &   d
\end{array}\right),
0 \leq d \leq 1.
\end{eqnarray}
It is nonunitary except for the identity operator $N(1)=I$. An especially interesting example of it is
$N(0)=
\left(
  \begin{array}{cc}
    1 & ~0\\
    0 & ~0\\
  \end{array}
\right)$,
describing a projective measurement $P_{0}=|0\rangle\langle0|$ in the computational basis.
Obviously, the diagonal matrix $D$ is equivalent to the quantum gate $N(d_{\theta})$ up to a real coefficient $d_{+}^{\theta}$, where the continuous variable $d_{\theta}=d_{-}^{\theta}/d_{+}^{\theta}=\sqrt{4-3\cos^{2}\theta}/(2+\sqrt{3}\cos\theta)$.

To proceed, the nonunitary gate $N(d_{\theta})$ can be described by $N(0)$ and one controlled-$U(d_{\theta})$ gate (denoted $\text{C}_{U(d_{\theta})}$) with the aid of an ancilla qubit initialized to $|0\rangle$,
where
\begin{eqnarray}
U(d_{\theta})=
\left(
  \begin{array}{cc}
d_{\theta}               &   \sqrt{1-d_{\theta}^{2}}\\
\sqrt{1-d_{\theta}^{2}}  &   -d_{\theta}
\end{array}
\right)
\end{eqnarray}
is a unitary gate. Actually, for an arbitrary single qubit system $a|0\rangle+b|1\rangle$, if one uses this system as control qubit and an ancilla qubit as target qubit, then the equality
\begin{eqnarray}
[I \otimes N(0)]\text{C}_{U(d)}[(a|0\rangle+b|1\rangle)|0\rangle] \equiv N(d)(a|0\rangle+b|1\rangle)
\end{eqnarray}
holds.
By this, one can realize the nonunitary diagonal matrix $D$ and thus the nonunitary evolution matrices $A_{i}$s.

\subsection{Success probability}

In the process of realizing the single qubit nonunitary gate, we apply the projective measurement on ancilla qubit.
Observation of the ancilla qubit in the computational basis then yields $|0\rangle$ or $|1\rangle$, probabilistically.
For the successful case that one gets $|0\rangle$, the success probability of the nonunitary gate $N(d_{\theta})$ is thus $p=|a|^{2}+d_{\theta}^{2}|b|^{2}$ for any normalized state $a|0\rangle+b|1\rangle$.
Because a unitary operator can always preserve inner products between state vectors, the success probability for operations $A_{i}$ is therefore
\begin{eqnarray}
p(|\psi_{i}\rangle)=
||[I \otimes N(0)]\text{C}_{U(d_{\theta})}[(V_{i}^{\dag}N_{i}|\psi_{i}\rangle)|0\rangle]||^{2}
=N_{i}^{2}(2-\sqrt{3}\cos \theta).
\end{eqnarray}
Note that for $\theta=\pi/2$, related to the BSM, we have $N_{i} = 1/\sqrt{2}$, $U(d_{\theta})=I$ and the success probability reaches the maximum 1.
In addition, comparing with the expression (\ref{probability-c}) we see that the complex coefficient $|c_{i}|^{2}=N_{i}^{2}(2-\sqrt{3}\cos \theta)$.

As an example, suppose that Alice obtains the result $\left |e_{10}  \right \rangle$ after EJM on her qubits 1 and 2, leaving Bob's qubit 3 in state (\ref{psi10}). Once Bob has known Alice's result of the EJM, he may immediately recover the original state
by applying unitary transforms $V_{10}^{\dag}$, $U_{10}$ on qubit 3 and the intermediate $\text{C}_{U(d_{\theta})}$ followed by the projective measurement on ancilla qubit, i.e.
\begin{eqnarray}
\frac{
[U_{10} \otimes I][I \otimes N(0)]\text{C}_{U(d_{\theta})} [(V_{10}^{\dag}N_{10}|\psi_{10}\rangle)|0\rangle]}
{||[U_{10} \otimes I][I \otimes N(0)]\text{C}_{U(d_{\theta})} [(V_{10}^{\dag}N_{10}|\psi_{10}\rangle)|0\rangle]||^{2}}
=|\psi_{0}\rangle,
\end{eqnarray}
with the success probability
\begin{eqnarray}
\frac{2-\sqrt{3}\cos \theta}{2 + \cos \theta (| \alpha |^{2} - | \beta  |^{2}) + 2\cos \theta \text{Re} [(1-\text{i}) \alpha ^{\ast}\beta]}.
\end{eqnarray}

\begin{figure}[!ht]
\centerline{\includegraphics[width=6.5in]{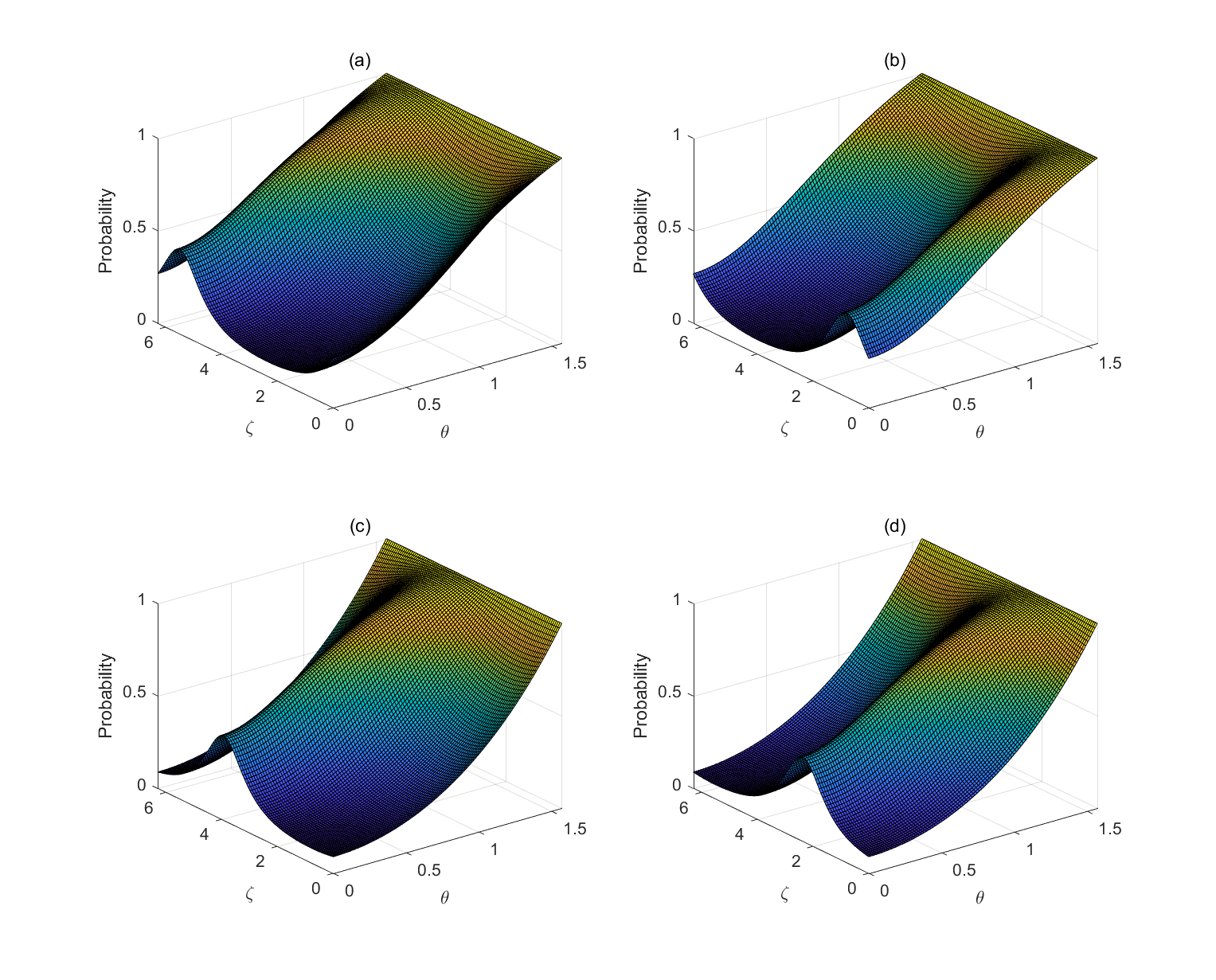}}
  \caption{The success probabilities of $A_{i}$s for teleporting an unknown single qubit state, $\cos (\zeta/2)|0\rangle+\sin (\zeta/2)|1\rangle$, $\zeta \in [0,2\pi]$, based on EJM parameterized by $\theta \in [0,\pi/2]$. For the nonunitary matrices (a) $A_{00}$, (b) $A_{01}$, (c) $A_{10}$ and (d) $A_{11}$, besides the maximum probability 1 for $\theta=\pi/2$, these success probabilities vary continuously over a range from $(2-\sqrt{3})/(2+\sqrt{2})$ to 1.}
  \label{probability.eps}
\end{figure}

Intuitively, let $\alpha=\cos (\zeta/2)$ and $\beta=\sin (\zeta/2)$, then we provide a plot of the probabilities in the range $0 \leq \zeta \leq 2\pi$ and $0 \leq \theta \leq \pi/2$, as shown schematically in Fig.\ref{probability.eps}.
It is straightforward to show that the maximum success probability 1 occurs at $\theta=\pi/2$, and in general, the success probabilities vary continuously over a range from $(2-\sqrt{3}\cos \theta)/(2+\sqrt{2}\cos \theta)$ to $(2-\sqrt{3}\cos \theta)/(2-\sqrt{2}\cos \theta)$.

Note that in EJM, Alice obtains the result $|e_{i}\rangle$ with $p_{i}$, probabilistically.
Together with the corresponding success probability of $A_{i}$, we now calculate the success probability for teleporting a single qubit state based on the EJM. We see that
\begin{eqnarray}
p=\sum_{i}p_{i} p(|\psi_{i}\rangle)=\sum_{i}\frac{1}{8N_{i}^{2}} \times N_{i}^{2}(2-\sqrt{3}\cos \theta)=1-\frac{\sqrt{3}}{2}\cos \theta.
\end{eqnarray}
Not surprisingly, for $\theta=\pi/2$, related to the BSM, the success probability is 1, and otherwise varies according to the parameter $\theta$.

\section{Quantum circuits}

\begin{figure}
\centerline{
\Qcircuit @C=1em @R=1em {
&\lstick{\text{qubit}~1: \ket{0}}  &\gate{R_{y}(\zeta)} &\gate{R_{z}(\xi)}  &\ctrl{1} &\gate{H}
&\ctrl{1}                           &\gate{S} &\gate{H}   &\meter    &\cctrl{2}_{~~m_{1}}\\
&\lstick{\text{qubit}~2: \ket{1}}              &\gate{H} &\ctrl{1} &\targ    &\qw &\gate{R_{z}(\frac{\pi}{2}-\theta)} &\gate{S} &\gate{H}   &\meter    &\cctrl{1}_{~~m_{2}}  \\
&\lstick{\text{qubit}~3: \ket{1}}              &\qw      &\targ    &\qw      &\qw
&\qw                                &\qw      &\qw        &\qw        &\gate{A_{{m_{1}}{m_{2}}}} &\qw \\
}}
\vskip 0.55\baselineskip
\centerline{\footnotesize}
\caption{Quantum circuit for quantum teleportation based on the EJM. The qubits 1 and 2 belongs to Alice, and qubit 3 to Bob. Each single line denotes a qubit and two double lines coming out of the measurement in computational basis carry classical bits $m_{1}$, $m_{2}$ for qubits 1 and 2, respectively.}
\label{circuit1}
\vskip 0.55\baselineskip
\end{figure}
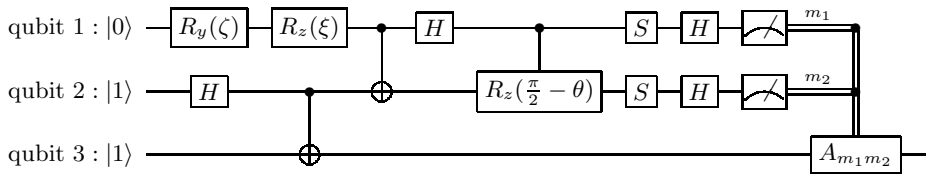

A quantum circuit for realizing the present scenario is shown in Fig.\ref{circuit1}.
It contains preparation of the initial state, EJM on Alice's system and the corresponding nonunitary quantum evolution by Bob.

The initial single qubit state can be easily prepared using two rotation operators about the $y$ and $z$ axes respectively, i.e. $R_{y}(\zeta)=\text{e}^{-\text{i}\zeta \sigma_{y}/2}$ and $R_{z}(\xi)=\text{e}^{\text{i}\xi/2}\text{e}^{-\text{i}\xi \sigma_{z}/2}$; the singlet state will be prepared using the Hadamard ($H$) gate $H=(\sigma_{x}+\sigma_{z})/\sqrt{2}$ and the controlled-NOT (CNOT) gate, where $\sigma_{x}$, $\sigma_{y}$ and $\sigma_{z}$ are respectively the Pauli sigma matrices \cite{NC2000}. The circuit for EJM \cite{TGB-EJM2021} successively consists of CNOT gate, $H$ gate, controlled-$R_{z}(\frac{\pi}{2}-\theta)$ gate
(a controlled phase shift gate, denoted $\text{C}_{R_{z}(\frac{\pi}{2}-\theta)}$), pairs of $S$ gates ($S=R_{z}(\pi/2)$) and $H$ gates followed by a measurement in computational basis, noted by $m_{1},m_{2}=0,1$, respectively for qubits 1 and 2.
This process can be described by
\begin{eqnarray}
(HS \otimes HS)\text{C}_{R_{z}(\frac{\pi}{2}-\theta)}(H \otimes I)\text{CNOT}|e_{i}\rangle =(-1)^{m_1 \oplus m_2}\text{i}|i\rangle,
\end{eqnarray}
where $m_1, m_2=0,1$ are two classical bits that provide the measurement results to Bob, and $i=m_1m_2=00,01,10,11$.

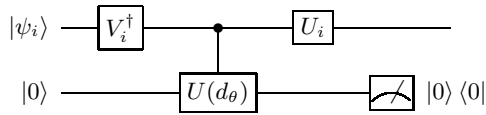
\begin{figure}
\centerline{
\Qcircuit @C=1.5em @R=1em {
&\lstick{\ket{\psi_{i}}}  & \gate{V_{i}^\dag} &\ctrl{1}    &\gate{U_{i}}  &\qw     &\qw\\
&\lstick{\ket{0}}         & \qw           &\gate{U(d_{\theta})} &\qw       &\meter  &~\ket{0}\bra{0}\\
}}
\vskip 0.55\baselineskip
\centerline{\footnotesize}
\caption{Quantum circuit for realizing the nonunitary matrices $A_{i}$s, $i=00,01,10,11$. The ancilla qubit (in bottom line) initiated in $|0\rangle$ is used to construct $\text{C}_{U(d_{\theta})}$ gate, and only the following result of $|0\rangle$ corresponds to the successful realization of the nonunitary operation.}
\label{circuit2}
\vskip 0.55\baselineskip
\end{figure}

The quantum circuit of nonunitary evolution matrix is shown in Fig.\ref{circuit2}, where $|\psi_{i}\rangle$ is the postmeasurement state of qubit 3 related to EJM-basis state $|e_{i}\rangle$ and the state $|0\rangle$ (in bottom line) is an ancilla qubit.
Two unitary operations $V_{i}^{\dag}$ and $U_{i}$, together with $\text{C}_{U(d_{\theta})}$ gate followed by the projection $|0\rangle\langle0|$ (i.e. $N(0)$) on the ancilla qubit, are capable of realizing the probabilistic nonunitary matrix $A_{i}$.

\section{Discussion and summary}

In summary, we have proposed a probabilistic teleportation scenario based on the EJM \cite{Gisin-EJM2019,TGB-EJM2021}, parameterized by $\theta \in [0,\pi/2]$.
Motivated by the idea that quantum entanglement enables new kinds of joint measurements \cite{Gisin-EJM2019}, the present work is the first attempt to develop quantum teleportation via this novel joint measurement.
There are a few interesting features in the present scenario.
First of all, it goes beyond the conventional teleportation scenario, which can be included in the present scenario. This is directly verified by taking $\theta=\pi/2$ in our scenario.
In fact, all of the teleportation scenarios based on the BSM can be extended to ones based on the EJM.
Second, it enables us to explore a kind of adjustable joint measurements involving multiple measurement settings.
Different from the well-known BSM being single input and four outcomes, the EJM can provide a continuously variable measurement settings for the sender (or the controller), which makes it truly useful---the ability to extend to port-based teleportation \cite{port-based-QT2008} or complex quantum communication networks \cite{Tavakoli-network2022}.
Meanwhile, we find that applying EJM causes undesired nonunitary quantum evolutions.
It is impossible to exactly implement a nonunitary quantum gate, but fortunately it can be realized probabilistically by the nonunitary quantum circuit \cite{TU-Nonunitary2005}.
To do so, we calculate the success probability of the present scenario. It has been shown that the success probability depends on phase factor in EJM, and undoubtedly the maximum probability of success 1 occurs at $\theta=\pi/2$, related to the BSM.

Moreover, we have shown in detail the quantum circuits to realize the present scenario.
The initial three-qubit state can be easily prepared conventionally, and also the circuit of EJM on Alice's system has been provided in \cite{TGB-EJM2021}. Thus, the most important requirement, in the present quantum circuits, is to be able to implement the nonunitary operations on Bob's qubit.
In view of this fact, we employ nonunitary gate (by means of an ancilla qubit and projection) and the accessible unitary gates to construct the circuits of nonunitary operations related to the outcomes of the EJM.
Notice that the reported experimental tests of nonbilocality networks, respectively in superconducting quantum computers \cite{BGT2021IBM} and hyperentangled photons \cite{PanPRL-EJM2022}, indicate the feasibility of the EJM.
The present scenario is experimentally feasible since only single qubit (or controlled) unitary gates and projective measurement are being utilized in the quantum teleportation circuits.

Finally, the present EJM-based method can be straightforwardly extended to multiqubit teleportation and we expect that this work will motivate further applications with the fascinating joint measurement.
Nevertheless, it remains some interesting open problems for EJM-based quantum information processing. For instance, one might be interested in how to construct a multiqubit EJM conditioned on preserving its elegant properties. Also, one needs to optimize and develop quantum circuit involving nonunitary gates, since its success probability decays exponentially with the number of nonunitary gates \cite{TU-Nonunitary2005}. In any event, the ultimate goal of quantum technology is to realize large-scale quantum information processing.

\begin{acknowledgements}
This work was supported by
the National Natural Science Foundation of China under Grant Nos: 62271189, 12071110,
the Hebei Central Guidance on Local Science and Technology Development Foundation of China under Grant No: 226Z0901G,
the Hebei 3-3-3 Fostering Talents Foundation of China under Grant No: A202101002,
the National Social Science Fund of China under Grant No: 23BZX103,
the Education Department of Hebei Province Natural Science Foundation of China under Grant No: ZD2021407,
the Education Department of Hebei Province Teaching Research Foundation of China under Grant No: 2021GJJG482.
\end{acknowledgements}

\end{document}